\documentclass[twocolumn,showpacs,preprintnumbers,amsmath,amssymb]{revtex4}
\begin{document}

\title{Dynamics of liquid $^4$He in confined geometries
from Time-Dependent Density Functional calculations}
\author{Luigi Giacomazzi, Flavio Toigo and Francesco Ancilotto}
\affiliation{Istituto Nazionale per la Fisica della Materia 
and Dipartimento di Fisica "G. Galilei", via Marzolo 8,
I-35131 Padova, Italy}
\date{\today}
\begin{abstract}
We present numerical results obtained from Time-Dependent Density
Functional calculations 
of the dynamics of liquid $^4$He 
in different environments 
characterized by geometrical confinement.
The time-dependent density profile and velocity 
field of $^4$He are obtained 
by means of direct numerical integration
of the non-linear Schrodinger equation associated
with a phenomenological energy functional 
which describes accurately both the
static and dynamic properties of bulk liquid $^4$He.
Our implementation allows for a general solution in 3-D 
(i.e. no symmetries are assumed in order to simplify the calculations).
We apply our method to study the real-time dynamics of 
pure and alkali-doped clusters,
of a monolayer film on a weakly attractive surface 
and a nano-droplet spreading on a solid surface.
\end{abstract}
\pacs{ 68.10.-m , 68.45.-v , 68.45.Gd }
\maketitle

\section{Introduction}
Density Functional (DF) methods 
\cite{evans} have become increasingly popular in recent years as 
a useful computational tool to study the properties
of classical and quantum inhomogeneous fluids, especially for
large systems where the DF methods provides a good compromise
between accuracy and computational cost.
In particular, a quite accurate description of the $T=0$ properties of
liquid $^{4}$He has been obtained within a DF approach by using the energy
functional proposed in Ref.\cite{dupont} and later improved in Ref.\cite
{prica}. 
In general, the Density Functionals for $^4$He depend on a number of
phenomenological parameters which are adjusted to
reproduce {\it bulk} experimental properties of liquid $^{4}$He at saturated
vapor pressure. 
The resulting functionals are found to describe accurately also
the properties of the inhomogeneous systems (e.g. the 
liquid-vapor interface of the free $^{4}$He surface)
(for a thorough comparison between
these and other functionals used to describe liquid $^{4}$He at $T=0$, see
Ref.\cite{szybis}). 

This phenomenological functional has been widely used in a
variety of problems involving inhomogeneous $^{4}$He systems
like pure and doped clusters \cite{dalfovo2,BarHer,dalfovo3},
electron "bubbles" in bulk $^4$He\cite{anci},
alkali atom adsorption on
the surface of liquid $^4$He \cite{anci1}, 
vortices in $^4$He clusters\cite{barranco1},
adsorption of liquid $^4$He on solid surfaces\cite{sartori,faccin}, etc.

According to Ref.\cite{prica} the static properties 
of liquid $^4$He at T=0 are described by the
following energy density functional 
(atomic units will be used throughout our paper):
\begin{widetext}
\begin{equation}\label{StatFunc}
E_0[\rho ]=\frac{1}{2 M}
\int d{\bf r} \left( \nabla
\sqrt{\rho}\right)^2
+\frac{1}{2}\int d{\bf r} \int d{\bf r'}\rho({\bf r})\rho({\bf r'})
V_{He-He}(|{\bf r}-{\bf r'}|)
+\int d{\bf r} \,E_c({\bf r})
\end{equation}
\end{widetext}
The first term on the right hand side 
is the quantum pressure (corresponding
to the kinetic energy of a Bose gas of nonuniform density). 
The second term contains a two-body pair potential
$V_{He-He}$ screened at short distances, while the last term
accounts for correlation effects due to the short-range part
of the He-He interaction.
The explicit form of the functional is given in Ref.\cite{prica},
to which the reader is referred for more details
(see also the Appendix in the present paper).

The minimization of the energy functional (\ref{StatFunc}) with 
respect to density variations,
subject to the constraint  of a constant number of $^4$He atoms,
$ \int \rho ({\bf r}) d{\bf r}=N $,
leads to the equilibrium profile $\rho ({\bf r})$,
thus allowing to study the {\it static} properties 
of the $^4$He system.

In order to study the {\it dynamics} 
of $^4$He 
an additional term (which has no effect on the static
properties derived from the minimization of the 
energy functional (1))
has been added \cite{prica} to the functional (1)  
in the form of a 
phenomenological current term: 
\begin{widetext}
\begin{equation}
H_J=\frac{\rho({\bf r})}{2}M{\bf v}({\bf r})^2
-\frac{M}{4}\int
d{\bf r}'V_{J}(|{\bf r}-{\bf r}'|)\rho({\bf r})\rho({\bf r}')
[{\bf v}({\bf r})-{\bf v}({\bf r}')]^2
\end{equation}
\end{widetext}

The first term is the usual hydrodynamic current density, while
the second term accounts in a phenomenological way for
non-local effects due to the 
"backflow" current density\cite{prica}.
The resulting DF (the so-called Orsay-Trento 
Functional), which will be used in our calculations,
has thus the following form:
\begin{equation}
E[\rho,{\bf v}]=E_0[\rho ]
+\int d{\bf r}\, H_J 
\end{equation}

One appealing feature of this functional,
which will prove to be essential in the time-dependent
calculations presented in the following Sections, is that it 
reproduces not only the static structure factor 
but also the bulk dispersion relations
of sound excitations in liquid $^4$He, i.e. the effective 
current-current interaction $V_J$ in (2) 
is empirically adjusted in order to reproduce the
phonon-roton dispersion curve of bulk $^4$He 
and the dynamic structure function as well.

The extended functional (3) 
has been applied in the past to the study 
of density oscillations in $^4$He 
clusters\cite{casas}, of surface excitations in $^4$He films   
\cite{prica,dalfovo2}
and of other related dynamical phenomena \cite{dalfovo}.
In all of the above applications a linear response
approximation was used, as well as suitable symmetries
were assumed in order to make the calculations feasible.

There are many different situations, however, where a
more general approach to 
investigate the microscopic dynamics of liquid $^4$He
in arbitrary environments, often characterized
by geometrical confinement, is highly desirable.
For instance, modifications of the excitation 
spectra due to geometrical
confinement are expected  
in the case of $^4$He films adsorbed on solid surfaces.
A second example involves impurity-doped $^4$He clusters.
Atomic impurities are currently used 
as experimental probes of the superconducting
behavior of $^4$He clusters: the 
coupling of the impurity motion with the cluster dynamics is 
crucial for a comprehension
of the observed spectra.
As a third example, we mention the problem of 
the dynamics of wetting phenomena: 
weakly attractive surfaces 
like the heavy alkali metal surfaces
are not wet by liquid $^4$He at very low temperature, and thus
finite amounts of $^4$He on these surfaces
will form a droplet characterized by a non-zero contact angle.
Even on a highly 
uniform surface, however, the contact angle between the
$^4$He droplet and the surface
is extremely hysteretic
and its value strongly depends on 
whether the contact line is advancing
or receding.  
Informations from microscopic calculations 
of the dynamic behavior of a liquid droplet/film spreading
on a weak solid surface can represent a valuable help to 
understand these effects.

We present here calculations, done in the framework
of Time-Dependent Density Functional (TD-DFT) phenomenological 
theory, of the dynamical properties of $^4$He (at T=0) in confined
geometries.  
Our methods allows to calculate the direct real-time
evolution of a $^4$He system in arbitrary $3-D$ geometries.
Some applications of the method, showing its efficiency and
capabilities, will be presented in the following Sections.

\section{Time-Dependent Density Functional Theory}

In order to extend 
the DFT scheme described in the previous Section to
the domain of time evolution
(Time Dependent Density Functional Theory, TD-DFT), 
one can use a variational principle as follows.
One starts from a suitable action integral $A$
written in terms of a complex "wave function"
$\Psi ({\bf r},t)$
\begin{equation}
A[\Psi]=\int dt d{\bf r} \left\{\mathcal{E}
[\Psi ,\Psi ^*]
-i\Psi ^*   {\frac{\partial \Psi }{\partial t}} \right\}
\end{equation}
where the energy density $\mathcal{E}$ is defined by
$E[\rho ,{\bf v}]=\int d{\bf r}\mathcal{E}$. 
From the condition of stationarity
\begin{equation}
 {\frac{\delta A}{\delta \Psi^*}}=0
\end{equation}
a time-dependent Euler-Lagrange equation follows:  
\begin{equation}\label{DynEq}
\left[-\frac{\nabla^2}{2M}+U[\rho,{\bf v}]
\right]\Psi\equiv H\Psi=i {\frac{\partial \Psi}{\partial t}} 
\end{equation}
The "effective potential" $U$ appearing in the above 
non-linear time-dependent Schrodinger equation
is defined as the variational derivative of the
energy functional, 
$U[\rho ,{\bf v}]\equiv \delta E[\rho ({\bf r})]/\delta \Psi ^*$
and its explicit expression is given in the Appendix. 
From the solution $\Psi\equiv \phi e^{i\Theta }$
of Eq.(\ref{DynEq}) one can get the time-dependent 
density $\rho({\bf r},t)=\phi^2$ and the fluid velocity field
${\bf v}({\bf r},t)=\frac{1}{M}\nabla \Theta$.

We describe in the following Section the numerical procedure
that we used in order to integrate numerically   
the Eq.(\ref{DynEq})\cite{nota}.

\section{Computational scheme}

The numerical solution of eq.(\ref{DynEq}), i.e.
the wave function at an arbitrary time $t$, $\Psi({\bf r},t)$, 
is obtained by using
the Crank-Nicholson's scheme (CN), which enables to evolve $\Psi(t)$
to a later time $t+\Delta t$:
\begin{equation}
\left[1+i\frac{\Delta t}{2}H
(t+\Delta t)\right]\Psi(t+\Delta t)=\left[1-i\frac{\Delta
t}{2}H(t)\right]\Psi(t)
\end{equation}
The CN algorithm is one of the many approximate methods 
that can be used to integrate numerically 
a time-dependent Schrodinger equation 
like the one shown in Eq.(\ref{DynEq}).
It is $2^{nd}$ order accurate in $\Delta t$, as can be easily seen 
by expanding the two sides of Eq.(7) in powers of $\Delta t$.
Among the advantages of such scheme is its
numerical stability for long-time evolutions, and the unitariness
(i.e. the number of particles is conserved during the
time evolution) \cite{AC}.
The CN recursion can be easily solved 
iteratively for each time step evolution. We find that three
iterations are usually enough to guarantee a stable
evolution even during long simulation times.

The application of the CN formula requires the calculation 
of the action of the
"Hamiltonian" $H\equiv -\frac{1}{2M}\nabla ^2 +U[\rho ,{\bf v}]$ 
to the complex wavefunction
$\Psi $.
A direct evaluation of the potential term $U$ 
(whose explicit expression is reported in the Appendix) would
be prohibitively costly, due to the presence of
convolution integrals of the form 
$I({\bf r})=\int d{\bf r}'f({\bf r}-{\bf r}')g({\bf r}')$,
whose evaluation in real space requires
a huge number of floating point operations.
For such reason we decided to employ a mixed real-reciprocal space
representation of the basic quantities $\Psi $ and of the
density $\rho $. 
Since the Fourier transform of a convolution is the product 
of the Fourier transforms of the two terms entering the integrals, 
one can perform efficiently the calculations of a convolution 
integral on a
computer by using Fast Fourier Transforms (FFT).
The use of FFT implies Periodic Boundary Conditions (PBC),
and thus our calculations have been performed 
using a periodically repeated
supercell containing $N$ Helium atoms.
Both $\Psi $ and $\rho $ 
are expanded in plane waves:
$\Psi ({\bf r})=\sum _{\bf G}
\Psi_{\bf G} e^{i{\bf G}\cdot {\bf r}}$,
and similarly for
$\rho $.
The ${\bf G}$'s are the supercell wave-vectors,
${\bf G}= \pi(n/L_x,m/L_y,p/L_z)$, with $n,m,p=0,\pm1,\pm2,...$
and $L_x,L_y,L_z$ are the sides of the supercell
(a primitive orthorhombic supercell is used in all our calculations).
The number of plane waves in the above expansions
is chosen 
such as to give converged values for the total energy 
$E$ and for the structural parameters of the 
$^4$He system under investigation.
In the various applications that we report in the following,  
the supercell size was always chosen in such a way to make
negligible the interaction of the $^4$He system with its repeated
images.

The action of the 
kinetic energy operator $\hat{T}\equiv -\frac{1}{2M}\nabla ^2$ 
on the wavefunction $\Psi $ is also evaluated in reciprocal space, i.e.
\begin{equation}
\hat{T}\Psi =\frac{1}{2M}FFT\{G^2 \Psi _G \}
\label{eq:lapla}
\end{equation}

Finally, the real-space expression of $H\Psi $, computed as
described above, is used in the 
iteration formula of the CN scheme to give the evolved
wave-function $\Psi (t+\Delta t)$.
A new potential $U$ is thus computed from the 
updated density and velocity, and the procedure is repeated for the 
next time-step.
Typical time steps $\Delta t$ which are found to give
stable time evolutions are of the order of 1-3 femtoseconds.

We stress the fact that no geometric or simmetry 
constraints are imposed during the time evolution
to reduce the computational effort.
Our computational scheme is thus general and can be applied
in particular to those situations characterized
by geometrical confinement of $^4$He. We will give in the 
following Sections few examples of such case.

\section{$^4$He bulk excitations}

One important feature of the most recent implementation
of DF for $^4$He used here is the 
inclusion of a "backflow" current term 
(the second term in Equation (2))
explicitly depending on the fluid velocity 
as well as on the fluid density,
such that the resulting functional is able to correctly reproduce
the bulk excitation spectrum of liquid $^4$He.
This can be immediately verified within our computational scheme
as follows.

We setup an initial state corresponding to a modulation of the 
bulk uniform density with a wave vector $\bf{q}$:
\begin{equation}\label{BulkExc}
\Psi({\bf r},t=0)=\left\{\rho_0[1+\epsilon \,sin({\bf q}
\cdot {\bf r})]\right\}^{1/2}
\end{equation}
where $\epsilon $ is a small amplitude and 
$\rho _0=0.02184\,\AA^{-3}$ is the saturation density
at zero temperature and pressure.
The system is then allowed to evolve in time
according to the CN scheme described above.
By monitoring the
periodic oscillations of  $\rho({\bf r},t)$ during the simulation, we
can immediately calculate the mode frequency:
\begin{equation}
\omega=\frac{2\pi}{T}=\omega({\bf q}).
\end{equation}
Here $T$ is the period of the observed oscillation.
By varying the wave vector ${\bf q}$ we get the 
whole dispersion curve of $^4$He, which
we compare in Figure 1 with the experimental data.
As it appears from Fig.1, the agreement 
with the measured data is very good in 
a wide range of values of the wavevector $q$.

We finally note that if the bulk dispersion relation is calculated
by neglecting the "backflow" term in eq.(2),
results deviating by as much as 50$\%$ from the experimental
curva are obtained for $q>0.5\,\AA ^{-1}$.

\section{Dynamics of $^4$He clusters}

The dynamics of liquid $^4$He clusters have 
been studied theoretically
in recent years by means of a number of 
different methods and thus
represents a convenient benchmark for our numerical scheme.

We have considered the {\it monopole} and {\it quadrupole}
vibrational modes of cluster of different sizes,
characterized by the angular momentum quantum number $l=0$ and
$l=2$, respectively.
The dipole mode ($l=1$) corresponds to a 
translation of the cluster as a whole,
and should thus lie at zero energy.
To study the monopole mode, which is 
a radial breathing oscillation of the cluster density,
we compute the main frequencies by monitoring the 
periodic variations with time 
of the mean square radius of the cluster:
\begin{equation}
R(t)=\sqrt{\frac{1}{N} {\int r^2 \rho ({\bf r},t) d {\bf r}}} .
\end{equation}

To study the quadrupole oscillations we analyze instead the 
time evolution of the quadrupole moment tensor $Q_{ij}$:
\begin{equation}
Q_{ij}(t)=\int \rho({\bf r},t)[3r_i r_j-r^2 \delta_{ij}] d^3 r.
\end{equation}
Suitable initial states for both modes 
are constructed by applying 
to the ground state wave function $\Psi_0=\sqrt{\rho _{eq}}$ 
(where $\rho _{eq}$ represents the equilibrium density profile for the
cluster, calculated as described in Section 1 by direct
minimization of the energy functional (1)) an
excitation operator $e^{i\xi F}$, where $F=r^2$
to excite monopole oscillations, while $F=r^2 Y_{20}$
to excite quadrupole oscillations:
\begin{equation}
\Psi({\bf r},t=0)= e^{i\xi F}\Psi_0({\bf r}).
\end{equation}
This initial state is thus allowed to evolve according to the
equation (6).
From a frequency analysis of the calculated
$M(t)$ and of $Q_{ij}(t)$ we then calculate 
both the amplitudes 
and the frequencies of the
observed oscillations.

We report in Fig.2 the calculated frequencies
of the main oscillations observed 
in the discrete region of the spectrum (i.e. below
the evaporation threshold $ \omega <|\mu |$,
$\mu $ being the chemical potential of a single $^4$He atom).

For comparison, 
we show in the same Figure the results 
taken from Ref.\cite{CaStr}, obtained within
the Random Phase Approximation (RPA) and 
using a zero-range Density Functional to describe the 
$^4$He system. 
The approach
used in Ref.\cite{CaStr}, which neglects completely the
short-range He-He correlation effects, 
is accurate only when applied 
to the study of excitations 
involving variations in space
larger than the average distance between the
$^4$He atoms: this is not, however, a major limitation 
for the long-wavelength
modes like the ones investigated here
(external field proportional
to $r^lY_{l0}(\hat {\bf r})$.

In the same plot we compare also our calculated frequencies 
with those obtained from
more recent calculations done by using a finite-range
DF including non-local effects, very similar to the
one used here, and again within RPA\cite{BarHer}.
The results are not much different from 
those obtained using a zero-range functional\cite{CaStr},
showing that the 
effects of non-localities are not
important in this case due to the 
$q=0$ character of the multipole
excitations studied here.

Our calculated values, as it appears from Fig.2,
do not differ much from the
two previous theoretical results, 
even if they are obtained by explicitly including 
the "backflow" term (see Eq.(2)) in the $^4$He density functional.
The reason is that this term becomes important in the
region where the bulk dispersion relation deviates
from its phonon-like behavior, whereas 
in the present case only low-q excitations are considered.
In the next sections we will present 
cases where the inclusion of the backflow term is 
instead crucial in order to get accurate results.

\section{Excitations of a $^4$He monolayer adsorbed on a 
weakly attractive surface}

The study of the structure, growth and excitations of 
liquid $^4$He films
adsorbed on solid surfaces is an 
important field of current research
on superfluid properties.
Reliable theories of inhomogeneous 
liquid $^4$He may represent a useful complementary tool
for studying the properties 
of liquid $^4$He interacting with a solid surface\cite{Krot} .

Because of the extremely weak
He-He interaction, liquid $^{4}$He in contact with almost any
substrate spreads to form a continuous film 
over the surface, so that vapor
and substrate are never in contact.
A remarkable exception to this behavior is found when $^{4}$He is adsorbed
on heavy alkali metal surfaces: due to the large electron spill-out at these
surfaces, alkali metals provide the weakest adsorption potentials in nature
for He atoms. Based on this observation, it was suggested \cite{cheng2,saam}
that $^{4}$He might not wet some heavy alkali metal surfaces. Subsequent
experiments have confirmed this remarkable prediction
\cite {rutledge,nacher,ketola,klier}.

We study here the excitations of 
a single {\it monolayer} of liquid $^4$He, adsorbed on 
a weakly attractive surface. We choose here the
surface of Rb:
this is because Rb is 
the weakest surface that it is actually wet by $^4$He at $T\sim 0$,
i.e. the stable state of liquid $^4$He on Rb is represented
by a thin liquid film covering the whole surface, rather
than being represented by a droplet as in the case of Cs.
The extremely weak He-surface interaction on the other hand
is expected to act as a small perturbation on the film.
Thus the excitations studied here should be representative 
to some extent of
those for an {\it isolated} $^4$He film.

To model the interaction with a Rb substrate, 
we use a binding potential $V_s(z)$ which 
describes the
interaction between Rb, occupying the half space $z\leq 0$,
and one $^4$He atom located at a distance $z$ above the surface,
which is taken to be ideally flat.
The explicit form for  $V_s(z)$, originally proposed 
in Ref.\cite{chizme}, has been later revised in Ref.\cite{faccin}
in order to correctly reproduce the experimental 
wetting properties of the $^4$He/Rb system.
We do not include any corrugation of the surface on the atomic 
scale to mimic its actual microscopic structure.
This is a good approximation for the case of $^4$He adsorption
on alkali metal surfaces, because the experiments \cite{diehl}  
indeed indicate that a surprisingly smooth surface 
is seen by adsorbed $^4$He atoms.

The total energy functional of liquid $^4$He interacting
with a Rb surface is thus described by the energy 
functional (3) augmented by the additional term 
$\int d{\bf r}\rho ({\bf r}) V_s({\bf r})$, which
describes the interaction energy due to the presence of the
surface.
We have preliminarly studied the equilibrium density 
of liquid $^4$He on the Rb surface, at different areal
coverages. Our results are reported in Fig.3,
where the density profiles in the direction perpendicular
to the surface (which is located at $z=0$) are shown. 
Note that as the coverage increases, a second layer starts
to form above the first layer. 
Quite arbitrarily, we take as
representative of a "monolayer" the profile shown
in Fig.3 with a thicker line, 
corresponding to $n=4.2\times 10^{-2}\,\AA ^{-2}$.

We thus prepare our monolayer in a suitable non-equilibrium
initial state, characterized by a density modulation
with a wave-vector ${\bf q}$ parallel to the surface.
We note that, due to the presence of a symmetry-breaking 
external potential (the He-surface potential)
which depends on $z$ (the coordinate normal to the
surface), such initial state will
be immediately coupled, during its time-evolution,
to excitations of the film polarized perpendicularly 
to the surface plane.

From a Fourier analysis of the density
variations during the time evolution
from this initial state, 
we are able to get the main excitations
of the $^4$He film.
Fig.4 shows our results. The filled and empty 
squares show the main features observed in the 
frequency spectrum of the time-dependent density
$\rho ({\bf r},t)$. The filled squares indicate the most
intense peaks, whereas the open squares indicate 
the somewhat weaker features in the spectrum.

For comparison we also show with solid line the
experimental "ripplon" dispersion relation\cite{lauter1}.
It appears from Fig.4 that at low $q$ values the 
lower disperion curve becomes softer that the ripplon
curve. This is not surprising
since the ripplon dispersion, which relates to the
surface excitations of a infinite half-space 
occupied by the liquid,
is altered by the finite thickness of the
film. A simple model calculation \cite{kala} 
gives in fact for the oscillation frequency of a thin 
$^4$He film of thickness $a$ in the low-$q$ limit:

\begin{equation}
\omega (q)\sim \omega _{ripplon}(q) (qa)^{1/2}
\label{eq:disp_film}
\end{equation}

Since $\omega _{ripplon} \sim q^{3/2}$, the expected
dispersion scales as $q ^2$, in 
qualitative agreement with our findings.
Although 
a definite character cannot be assigned unambiguously 
to the observed mode shown in Fig.4
due to the strong mixing between the two bands,
we suggest that the upper branch should have
a strong component of density modulations
perpendicular to the surface.
To verify this we have 
induced in the monolayer a simple $q=0$ 
mode where the $^4$He film 
is initially rigidly shifted (by a very small amount)
towards the surface.
Due to the presence of the holding He-surface
potential, the film once let free to evolve, 
execute a motion where it executes oscillations mainly 
perpendicul to the surface, with
a main frequency at about 5 K: this value is shown 
with a filled dot at $q=0$ in Fig.4.

We can compare our results for the film 
excitations 
with those obtained in Ref.\cite{clements1}
by means of a microscopic statistical theory, where
the dynamic excitations of {\it thick} $^4$He films on a Cs surface
were considered.
In that work it is found that ripplon modes appear
in the calculated spectra, whose
dispersion follows the $q^{3/2}$ law for surface
excitations, as well as "interfacial" ripplons localized
mainly at the $^4$He/surface interface.
In particular, the lowest mode calculated  
in Ref.\cite{clements1} (corresponding 
the the lower branch in Fig.4)
has a ripplon character.
The second mode has a "longitudinal" character
(i.e. parallel to the surface)
on stronger substrates, 
only on very weak surfaces (i.e. Cs) it has a 
visible perpendicular component.
A linear (third sound) dispersion 
has been also observed in Ref.\cite{clements1}
as long as the wavelength is large compared
with the film thickness. We do not observe any such modes, 
probably because of the finite thickness of our
system.

We believe that the upper branch in Fig.4 
is the counterpart of a similar
excitations observed in 
neutron diffraction experiments on $^4$He adsorbed
on a graphite substrate\cite{lauter}, 
where, among other excitations, 
almost dispersionless 
modes at low frequencies have been observed.
These mode have been interpreted \cite{Apaja},
by means of microscopic calculations, 
as standing waves of the $^4$He liquid perpendicular to the 
substrate. 
We cannot compare directly 
our results with
the experimental measurements of \cite{lauter}
and with the microscopic calculations of Ref.\cite{Apaja}
because of the different substrate used 
(graphite is a much "stronger" substrate than alkali
metal surfaces)
and because multylayer films were considered
in these references.
We hope that our results will stimulate further
experimental measurements, in particular to study 
the excitations of $^4$He monolayers
on weakly attractive surfaces.

\section{Dynamics of alkali-doped $^4$He clusters}

The spectroscopy of doped liquid $^4$He clusters has become a 
current tool for
the study of superfluidity in $^4$He droplets\cite{toennies}.
As a consequence of the ultra-weak alkali-He interaction,
an alkali atom picked up by 
a $^4$He cluster has its stable state in a "dimple" on the
surface of the cluster\cite{anci1,ernst1}, rather than being solvated into
its interior like most of the other atomic impurities.
Thus alkali could provide an excellent probe
for the {\it surface} excitation of superfluid
nanodroplets\cite{ernst1}.

The analysis and interpretation of
experimental studies of the formation of 
$^4$He-alkali complexes 
in doped clusters
would benefit from theoretical work of the excitations
of impurities attached to $^4$He clusters.

Although for light alkali the motion of the impurity
can be considered to a first approximation as occurring
in a {\it static} $^4$He environment (i.e. He does not adjust
appreciably to the istantaneous atom position),
however, for heavy alkali (Rb,Cs) a full dynamical approach
is required, where the $^4$He atoms in the cluster are allowed 
to adjust
dynamically upon the motion of the adsorbed impurity.

We address here the problem of finding the spectrum 
of vibrational excitations of a "dimple" state.
We consider in particular the system composed of 
a single Rb atom (in its electronic ground-state) 
attached to a 300-atom $^4$He cluster.
The Rb-He interaction is taken from Ref.\cite{patil}.

Fig. 5 shows the calculated equilibrium density profile 
(in a plane containing the center of the cluster and
the Rb atom) for the "dimple"
state representing the lowest energy configuration
of the Rb-$^4$He system.

By applying to the Rb atom a small 
initial momentum towards the 
surface of the cluster, we observe that the impurity
starts oscillating around its equilibrium position
shown in Fig. 5, and at the same time 
the $^4$He density changes in time
to adjust dynamically to the istantaneous Rb position.
Our results for the Rb dynamics are shown in Fig.6.
In the upper panel we show the calculated istantaneous 
position of the Rb atom as a function of time (measured
with respect to the Rb-cluster center-of-mass).
A straightforward Fourier transform of the signal
shown in the upper panel 
allows to compute the frequency 
spectrum shown in the lower panel.

We compare our results with a previous 
estimate \cite{ernst2} based on {\it static} DF calculations,
where the frequency of the oscillations of a Rb atom
attached to a 300-atom cluster was estimated approximately by assuming 
that the He atoms adiabatically adjusted to the istantaneous 
position of the impurity.
In that case the value of the frequency was
extracted from the shape of the impurity-He 
potential energy surface, giving $\omega \sim 0.7\,K$.
It appears from our results that a more complex 
dynamics results, due to the coupling of the Rb
atom motion with the surface oscillations of the 
cluster.
The frequency of a surface mode of an N-atom
$^4$He cluster can be written as
$
\omega = \sqrt{l(l-1)(l+2)}\omega _0/\sqrt{N}
$
where $\omega_0=3.45\,K$.
This is approximately $\sim 0.6\,K$ 
for the lowest $l=2$ mode, i.e. close to the 
natural Rb frequency calculated in Ref.\cite{ernst2}. 
A large amount of mixing is thus expected, which
determines the appearance of the complex vibrational
spectrum shown in Fig.6.

Experimental measurements of such oscillations
are under way \cite{ernst_private} to confirm our predictions.
A more complex and challenging application of
time-dependent methods, i.e. the
study of the oscillations 
of an alkali impurity in an {\it excited} electronic state,
is currently in progress.

\section{Spreading of a $^4$He nanodroplet 
on a weakly attractive substrate}

The mechanism underlying the spreading of a 
liquid droplet on a solid substrate
is a long-standing problem. Experiments done on a number 
of differente (classical) molecular fluids
have revealed universal laws such as the Tanner' spreading law
\cite{tanner}, where the droplet radius grows as
$r(t)\propto t^{1/10}$,
as well as fascinating phenomena on the atomistic scale (such as
the appearence of a monolayer 
film advancing in front of the drop\cite{heslot}, two different
regimes of the growth of the radius with time\cite{coninck1},
dynamical "layering", etc.).
Other experiments\cite{heslot} have shown that one or more monomolecular 
precursor layers spread ahead of the droplet cap 
with an average radius $R(t)\sim \sqrt{t}$.

Numerical studies of droplet spreading have been performed
\cite{blake,koplik,nieminen,coninck1} in recent years.
Despite many efforts, however,
many microscopic details of the spreading process
are however still to be known.

Even more challenging is the 
behavior of a superfluid $^4$He droplet
on a weakly attractive surface:
it has been found that $^4$He droplets 
adsorbed on a Cs substrate have spreading 
and flow properties that are not simple consequences 
of bulk superfluid behavior\cite{science}.
When a $^4$He droplet is deposited on a Cs surface 
(the only material known that is
not wetted by superfluid $^4$He at very low temperature), 
even on a highly 
uniform surface the contact angle between the
droplet and the surface
is extremely hysteretic
and its value strongly depends on 
whether the contact line is advancing
or receding.
Superfluid $^4$He droplets on Cs are also remarkable because they can
resist flow against a substantial 
chemical potential gradient\cite{science}.

We have studied the spreading of a 200-atom He cluster
on a weakly attractive surface. We choose for the latter
the case of Cs
(but very similar results are obtained for a Rb surface).
To simulate the Cs surface, we use the "ab initio" 
potential developed in Ref.\cite{chizme}, which has applied
successfully to the wetting properties of $^4$He an a Cs surface
\cite{faccin}.
The initial state is represented by a spherical cluster
(whose density profile is obtained by minimization of the
static functional (2) in the absence of the attractive 
He-surface potential), placed in the vicinity of the surface.
The downward force responsible for the spreading of the droplet
is provided solely by the long-range Van der Waals forces exerted by
the substrate
(due to the microscopic size of the droplet, 
the effect of gravity on spreading is negligible).

The spreading will continue, until the 
stable state is eventually reached.
The nature of the latter depends on the strength of
the He-surface potential: in the case of Cs, which is not wet
by He at T=0, the final configuration will be
that of an almost spherical cap characterized by a contact angle
of $\sim 40\,^\circ$ (partial wetting behavior)\cite{science,sartori}. 
More attractive substrate, on the other hand, 
will eventually be covered 
by a uniform microscopic 
film (complete wetting behavior).
Due to the finiteness of our systems and our use of 
periodic boundary conditions
our simulations are necessarily interrupted  
when the droplet hits its repeated images
and starts coalescing with them.

The sequence characterized the spreading process is shown
in Fig.7.
The overall duration of the simulation shown in the Figure is
about $2\,ns$.
Interestingly enough, a precursor layer seems to form
under the cap of the spreading cluster. We have computed 
from the cluster density profile $\rho ({\bf r},t)$ the 
average radius $R$ of this precursor layer.
The dependence of $R$ from the time, shown in Fig.8,
shows a "fast" regime where
the precursor layer expands linearly in time,
as found in similar simulations for 
classical fluids\cite{nieminen}.
Then a slowing down occurs, in correspondence of the formation of the
second layer (see panel (f) in Fig.7). 
It seems that after a transient, an almost linear 
spreading occurs again, although with 
a somewhat lower velocity.
The speed from the linear portion of Fig.8 gives a velocity
of about $\sim 50\,m/s$, which is close to the critical
Landau velocity. 
The slowing down is perhaps due to the spontaneous formation of 
vortical rings during the spreading. One such ring
is shown in Fig.9 by means of a vector plot showing the
current density ${\bf J}\equiv \rho (\bf r){\bf v}(\bf r)$
in a plane containing the center of the cluster and perpendicular
to the Cs surface.
Fig.9 also clearly shows the flow of atoms from the droplet cap
feeding the underlying precursor film.

Although transient precursor films spreading linearly with
time has been obtained from MD simulations of
classical fluids\cite{nieminen,blake}, they have not been
seen in any experiment.
This is probably due to the fact that they 
represent the very early stage of the spreading process:
as the equilibrium shape of the droplet 
is reached, the initial fast spreading is found to slow down
to a power law $R(t)\sim t^{1/7}$\cite{ruijter},
when a regime dominated by dissipation
due to the viscous flow starts to dominate.
It would be extremely interesting to study 
how this slowing down (which is only suggested 
from our short-time results
in Fig.8) occurs in the case of superfluid flow.
  
The presence of surface disorder  
is expected to alter the dynamics of 
liquid $^4$He films spreading
on a solid surface,
because the three-phase contact 
line of the spreading film/droplet
can easily be pinned to these defects, 
modifying the velocity of the contact line
as it advances (or recedes) on the surface.
We are currently investigating the effect of simple 
defects
on the atomic-scale (adatoms or vacancies) 
on the droplet spreading process 
discussed above.

\section{Summary}
We have presented TD-DFT calculations of the dynamical properties
of $^4$He in situations characterized by geometrical confinement.
We have shown the feasibility and efficiency 
of fully 3-D TD-DFT calculations
to study the dynamics of liquid He-4 in a number
of representative situations:
the dynamics of pure and alkali-doped $^4$He clusters,
the excitations of a liquid $^4$He monolayer adsorbed
on a weakly attractive surface and the dynamics of
a $^4$He nano-droplet spreading on a Cs surface.  

\bigskip

Acknowledgments:
We thank M.~Pi, M.~Barranco, E.~Krotscheck, W.~Ernst, 
F.~Dalfovo, V.A.~Apkarian and M.W.~Cole for useful 
comments and discussions.
We thank V.A. Apkarian for giving us a copy of the 
manuscript \cite{Ara} prior to publication.
We acknowledge funding from MIUR-COFIN 2001.

\bigskip
\bigskip
\bigskip
\centerline{\bf Appendix}

The effective potential entering Eq.(6) can be readily
evaluated by functional differentiation of the energy
functional (3) (see Ref.\cite{prica} for its detailed
expression):

\begin{widetext}
\begin{eqnarray}
U[\rho,{\bf v}]&=&V_s({\bf r})+\int d{\bf r'}\rho({\bf r}')V_{\ell}
(|{\bf r}-{\bf r}'|)+\frac{c_{2}}{2}
\bar{\rho}({\bf r})^2+\frac{c_{3}}{3}\bar{\rho}({\bf r})^3 
+\int d{\bf r'} \rho({\bf r}')[
c_{2}\Pi_{h}(|{\bf r}-{\bf r}'|)\bar{\rho}({\bf r}')+
c_{3}\Pi_{h}(|{\bf r}-{\bf r}'|)\bar{\rho}^2({\bf r}')]
\nonumber \\
& &+\frac{\alpha _{s}}{2M}\left(1-\frac{\rho({\bf r})}{\rho_0}\right) 
\int d{\bf r'}
\left(1-\frac{\rho({\bf r}')}{\rho_{0}}\right)
\nabla_{r'}\rho({\bf r}')
\cdot 
\nabla_{r} F(|{\bf r}'-{\bf r}|)
-\frac{M}{2}\int
d{\bf r'}V_{J}(|{\bf r}-{\bf r}'|)\rho({\bf r}',t)(v({\bf r})-v({\bf r}'))^2
\nonumber \\ & &+\frac{i}{2\rho({\bf r})}\nabla \cdot \int d {\bf r}' \, 
{\bf g} ({\bf r},{\bf r}') 
\end{eqnarray}
\end{widetext}

$V_s({\bf r})$ represents an additional external potential
acting on the $^4$He system.
The second term contains a two-body $^4$He-$^4$He pair potential
$V_{\ell}(r)$ screened at distances
shorter than a characteristic length $h_{\ell}$, while the third
and the fourth term (correlation terms),
which contains the average of the density
over a sphere of radius $h_{\ell}$, $\bar{\rho}_{r}$,
accounts for the internal kinetic
energy and for the increasing contribution of the hard-core
He-He repulsion when the density is increased.
The last two terms represent the contribution to the potential $U$ 
of the 
"backflow" term (2) 
introduced in Ref.\cite{prica}
in order to reproduce the bulk excitation spectrum of liquid $^4$He.
In particular, 
the last term contains the "backflow" current
${\bf J}_B \equiv \int d {\bf r}'
{\bf g} ({\bf r},{\bf r}')=\int d {\bf r}' V_{J}(|{\bf r}-{\bf r}'|)
\rho({\bf r})\rho({\bf r}')({\bf v}({\bf r})-{\bf v}({\bf r}'))$

The free parameters $h_{\ell}$, $c_2$, $c_3$, $\alpha _s$ are
adjusted in order to reproduce a number of experimental
properties of {\it bulk} liquid $^4$He.
For a detailed description of the various terms
and the numerical values of the adjustable parameters, 
we refer the interested reader to Ref.\cite{prica}.

\bigskip
\bigskip
\bigskip
\bigskip
\bigskip

\centerline{Figure Captions}

\bigskip
\bigskip

Figure 1: Bulk $^4$He excitation spectrum.
The points show the calculated spectrum, while the 
solid line shows the experimental dispersion relation.

\bigskip

Figure 2: Calculated monopole (upper lines) and quadrupole (lower lines)
oscillation frequencies of liquid $^4$He clusters, 
shown as a function of the
cluster sizes. Dots: this work; squares: results from
Ref.\cite{CaStr}; triangles: results from Ref.\cite{BarHer}.

\bigskip

Figure 3: Equilibrium density profiles for $^4$He films adsorbed on
a Rb surface. The z direction is perpendicular to the surface plane.
Different curves represent increasing
values of the surface coverages (number of $^4$He 
atoms per unit surface area),
from $n=1.6\times 10^{-2}\AA ^{-2}$ 
(lower profile)
to $n=7.2\times 10^{-2}\AA ^{-2}$ (upper profile).
The thicker line (defining our "monolayer") 
is drawn at $n=4.2\times 10^{-2}\AA ^{-2}$.

\bigskip

Figure 4: Calculated dispersion curves for a $^4$He monolayer
adsorbed on a Rb surface (squares). The dotted line is the
bulk dispersion relation, while the solid line shows the
experimental "ripplon" dispersion.

\bigskip

Figure 5: Density contour plot for a 300-atom $^4$He cluster
with a Rb atom attached on its surface: 
the dot shows the equilibrium
position of the Rb impurity.

\bigskip

Figure 6: Upper panel: Rb displacement with respect to the 
center-of mass of the Rb-cluster system, 
measured along the radial direction,
as a function of the simulation time.
Lower panel: Fourier spectrum (in arbitrary units) 
of the time series shown in the upper panel.

\bigskip

Figure 7: Sequence of $^4$He density contour plots 
showing the spreading of a 200-atom cluster on a Cs surface.

\bigskip

Figure 8: Radius of the precursor layer as a function of time.

\bigskip

Figure 9: Vector plot of the current density $\bf \rho \bf v $
in a plane perpendicular
to the surface and passing through the center of the cluster.
The configuration shown corresponds to the density contours in
panel (e) of Fig.7.


\bigskip
\bigskip
\bigskip


\begin{thebibliography}{99}
\bibitem{evans}  R.Evans, Adv. Phys. {\bf 28}, 144 (1979).
\bibitem{dupont}J.Dupont-Roc, M.Himbert, N.Pavloff and J.Treiner,
J. Low Temp. Phys. {\bf 81}, 31 (1990).
\bibitem{prica}F.Dalfovo, A.Lastri, L.Pricaupenko, S.Stringari and J.Treiner,
Phys. Rev. B {\bf 52}, 1193 (1995).
\bibitem{szybis}  L.Szybisz, Eur. Phys. J. B {\bf 14}, 733 (2000).
\bibitem{dalfovo2} F. Dalfovo, Z. Phys. D {\bf 29}, 61 (1994).
\bibitem{BarHer} M. Barranco, E.S. Hernandez,
Phys. Rev. B  {\bf 49}, 12078, (1994).
\bibitem{dalfovo3} F. Dalfovo, Phys. Rev. B {\bf 46}, 5482 (1992).
\bibitem{anci}F.Ancilotto and F.Toigo, Phys. Rev. B 50, 12820, (1994).
\bibitem{anci1}F.Ancilotto, E.Cheng, M.W.Cole and F.Toigo,
Z. Phys. B {\bf 98}, 323 (1995).
\bibitem{barranco1} F. Dalfovo, R. Mayol, M. Pi and M. Barranco,
Phys. Rev. Lett. {\bf 85}, 1028 (2000).
\bibitem{sartori}F. Ancilotto, A.M. Sartori and F. Toigo, Phys. Rev. B
{\bf 58}, 5085 (1998).
\bibitem{faccin} F.Ancilotto, F.Faccin and F.Toigo,
Phys. Rev. B 62, 17035 (2000).
\bibitem{casas}M.Casas, F. Dalfovo, A.Lastri, Ll.Serra, and S.Stringari,
Z. Phys. D {\bf 35}, 67 (1995).
\bibitem{dalfovo}F.Dalfovo, A.Franchetti,A.Lastri,L.Pitaevskii and S.Stringari,
Phys. Rev. Lett. {\bf 75}, 2510 (1995);J. Low Temp. Phys. {\bf 104}, 367
(1996).
\bibitem{nota}
We are aware of another 
computational scheme for TD-DFT
developed recently
\cite{Ara} to solve the equation (6),
which has been used to study
the dynamics of excess electron in liquid $^4$He "bubble" states.
At variance with the scheme proposed in Ref.\cite{Ara},
which is limited to radially symmetric dynamics,
our implementation does not assume any symmetry in the system.
\bibitem{Ara} J.~Eloranta and V.A.~Apkarian,
submitted for publication to J. Chem. Phys. (2002).
\bibitem{AC} A. Askar and A.S. Cakmak, J. Chem. Phys. {\bf 68}, 2794 (1978).
\bibitem{CaStr} M. Casas e S. Stringari, J. Low Temp. Physics {\bf 79}, 135
(1990).
\bibitem{Krot} B.E.Clements, H.Forbert, E. Krotscheck, H.J.Lauter,
M.Saarela and C.J. Tymczak, Phys. Rev. B {\bf 50}, 6958 (1994).
\bibitem{cheng2} E.Cheng, M.W.Cole, W.F.Saam and J.Treiner,
Phys. Rev. Lett. {\bf 67}, 1007 (1991).
\bibitem{saam}E.Cheng, M. W. Cole, W. F. Saam, and J.Treiner,
Phys. Rev. {\bf 46}, 13967 (1992).
\bibitem{rutledge}J.E.Rutledge and P.Taborek, Phys. Rev. Lett
{\bf 69}, 937 (1992).
\bibitem{nacher}P.J.Nacher and J.Dupont-Roc, Phys. Rev.Lett
{\bf 67},2966 (1991).
\bibitem{ketola}  K.S. Ketola, S. Wang, and R.B. Hallock, Phys. Rev. Lett.
{\bf 68} 201 (1992).
\bibitem{klier} J. Klier, P. Stefanyi and A.F.G. Wyatt, Phys. Rev. Lett.
{\bf 75}, 3709 (1995).
\bibitem{clements1} B.E.Clements, E. Krotscheck
and C.J. Tymczak, J. Low Temp. Phys. {\bf 107}, 387 (1997);
B.E.Clements, E. Krotscheck and M.Saarela, Z. Phys. B {\bf 94}, 115 (1994).
\bibitem{chizme}A. Chizmeshya, M.W.Cole and E. Zaremba,
J. Low Temp. Phys. {\bf 110}, 677 (1998).
\bibitem{diehl} J.D.White, J.Cui, M.Strauss, R.D. Diehl, F.Ancilotto and
F.Toigo, Surf. Sci. {\bf 307}, 1134 (1994).
\bibitem{lauter1}  H.J. Lauter, H. Godfrin, V.L.P. Frank and
P.Leiderer, Phys. Rev. Lett. {\bf 68}, 2484 (1992). 
\bibitem{kala}I.M.Khalatnikov, G.V. Kolmakov and V.L. Pokrovsky,
Soviet Physics JETP {\bf 80}, 873 (1995).
\bibitem{lauter} H.J. Lauter, H. Godfrin and H.Wiechert,
in {\it Proceedings of the Second International Conference on 
Phonon Physics}, J.Kollar et al. (eds.),
World Scientific, Singapore (1985), p.842.
\bibitem{Apaja} V.Apaja, H.Godfrin, E. Krotscheck and
H.J.Lauter, J. Low Temp. Phys. {\bf 124}, 599 (2001).
\bibitem{toennies} J.P.Toennies and A.F.Vilesov, 
Annu. Rev. Chem. {\bf 49}, 1 (1998); 
K.K.Lehmann and G.Scoles, Science {\bf 279}, 2065 (1998).
\bibitem{ernst1} F. Stienkemeier, W.E. Ernst, J.Higgins and
G.Scoles, J. Chem. Phys. {\bf 102}, 615 (1995).
\bibitem{patil} S.H.Patil, J. Chem. Phys. {\bf 94}, 8089 (1991).
\bibitem{ernst2} F.R.Bruhl, R.A.Trasca and W.E. Ernst, 
J. Chem. Phys. {\bf 115}, 10220 (2001).
\bibitem{ernst_private} W.E. Ernst, private communication. 
\bibitem{tanner}L.H.Tanner, J. Phys. D {\bf 12}, 1473 (1979).
\bibitem{heslot} F.Heslot, A.M.Cazabat and P.Levinson,
Phys. Rev. Lett. {\bf 62}, 1286 (1989).
\bibitem{coninck1} J.De Coninck,U.d'Ortona, J.Koplik and J.R.Banavar,
phys. Rev. Lett. {\bf 74}, 928 (1995).
\bibitem{blake} T.D.Blake, A.Clarke, J.De Coninck,M. de Ruijter and
M.Voue', Colloids and Surfaces {\bf 149}, 123 (1999).
\bibitem{koplik} J. Yang, J.Koplik and J.R.Banavar, Phys. Rev. Lett.
{\bf 67}, 3539 (1991).
\bibitem{nieminen} J.A. Nieminen, D.B.Abraham, M.Kartunnen and K.Kaski,
Phys. Rev. Lett. {\bf 69}, 124 (1992).
\bibitem{science}D.Ross, J.E.Rutledge, P.Taborek,
Science, {\bf 278}, 664 (1997).
\bibitem{ruijter} M.J.de Ruijter, J.De Coninck and G.Oshanin, Langmuir
{\bf 15}, 2209 (1999).
\end{thebibliography}
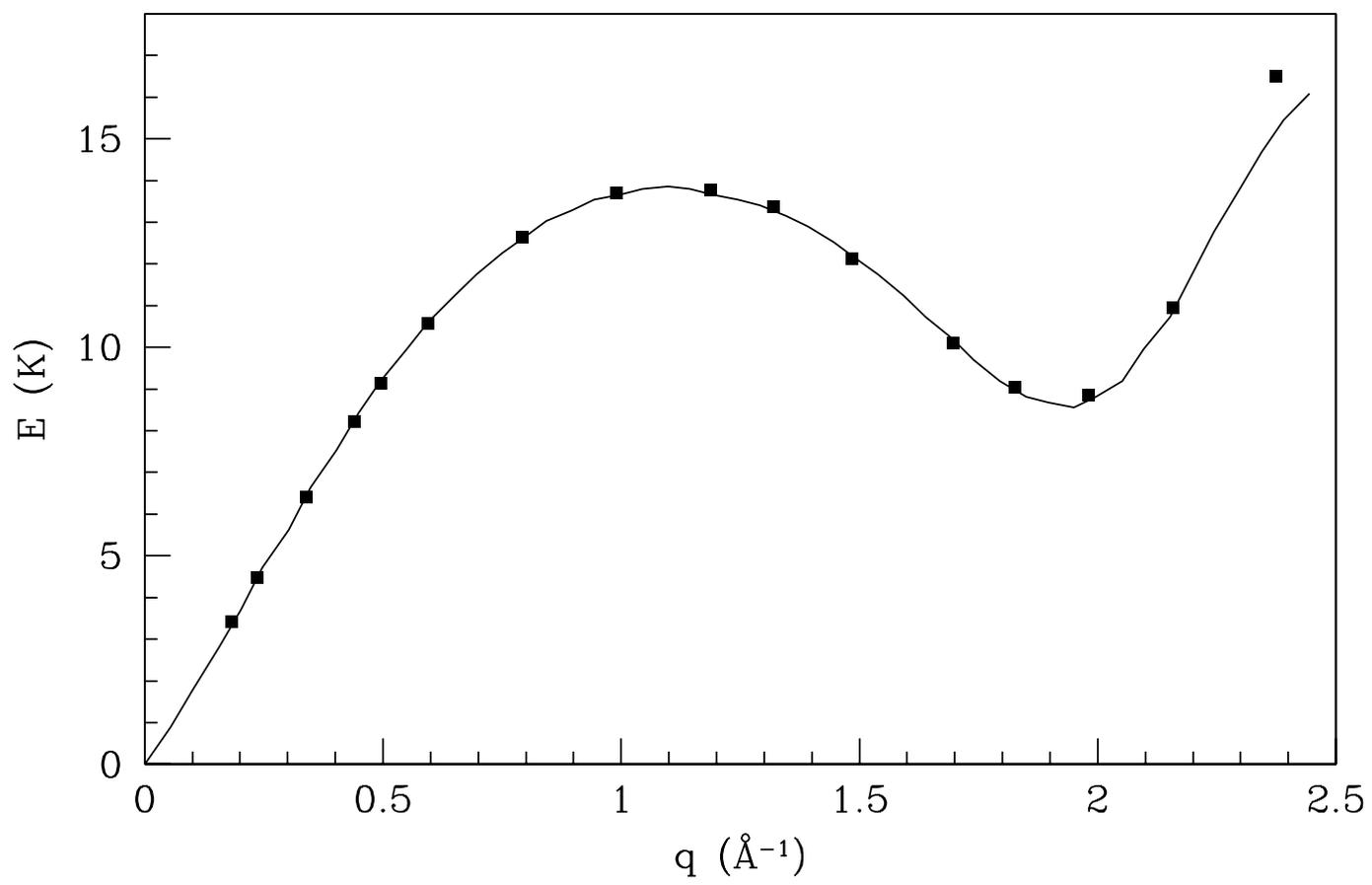
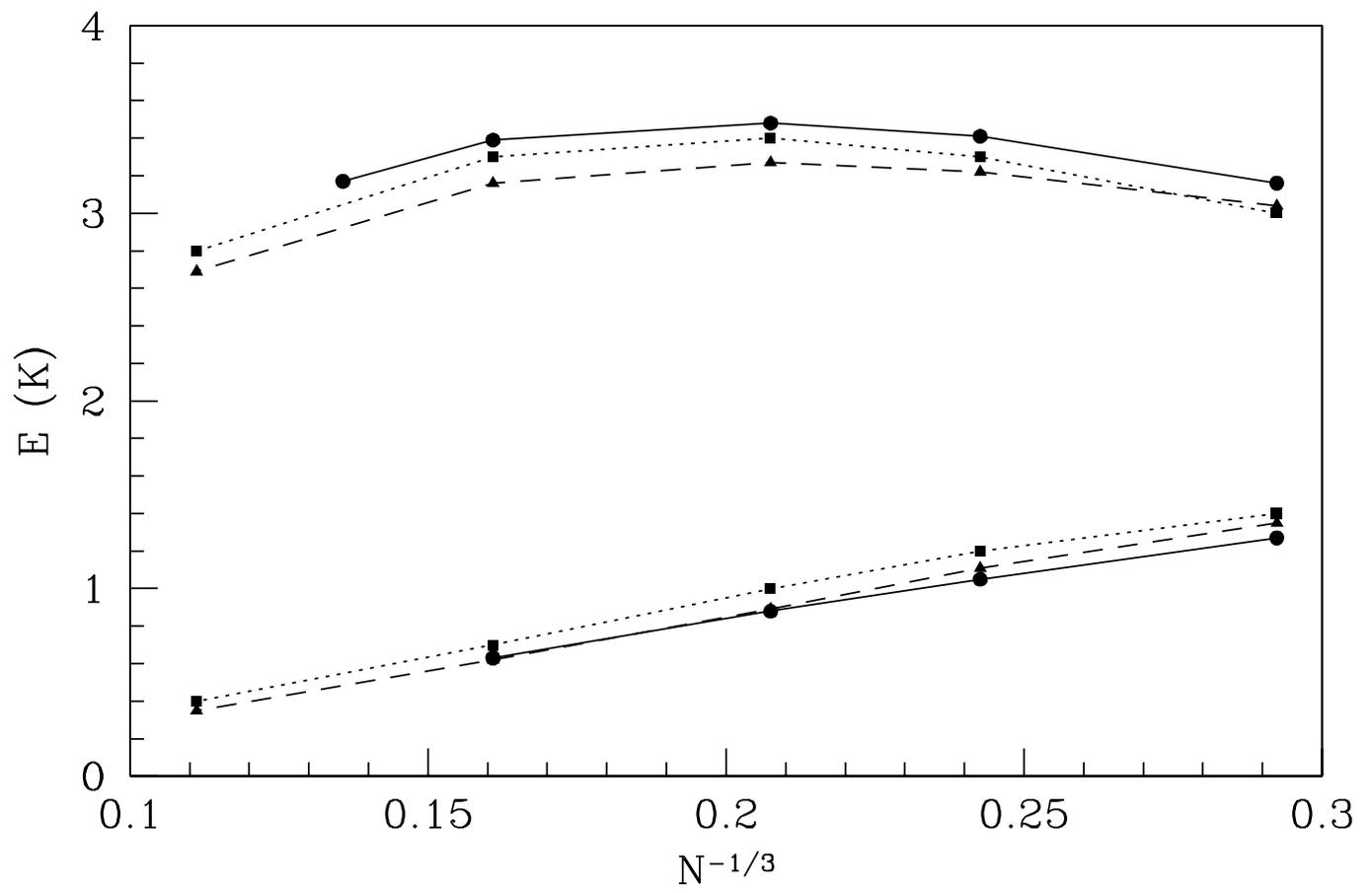
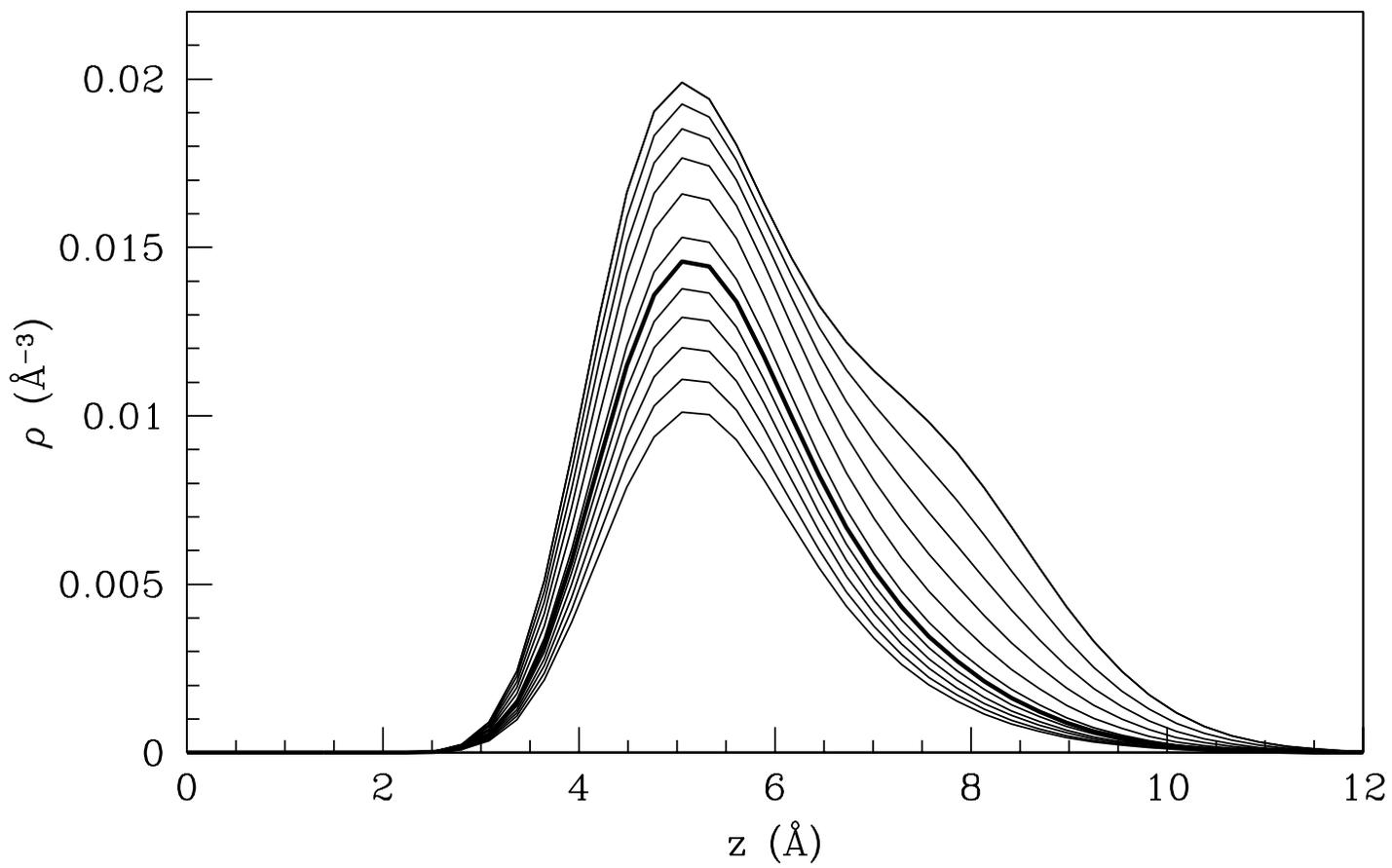
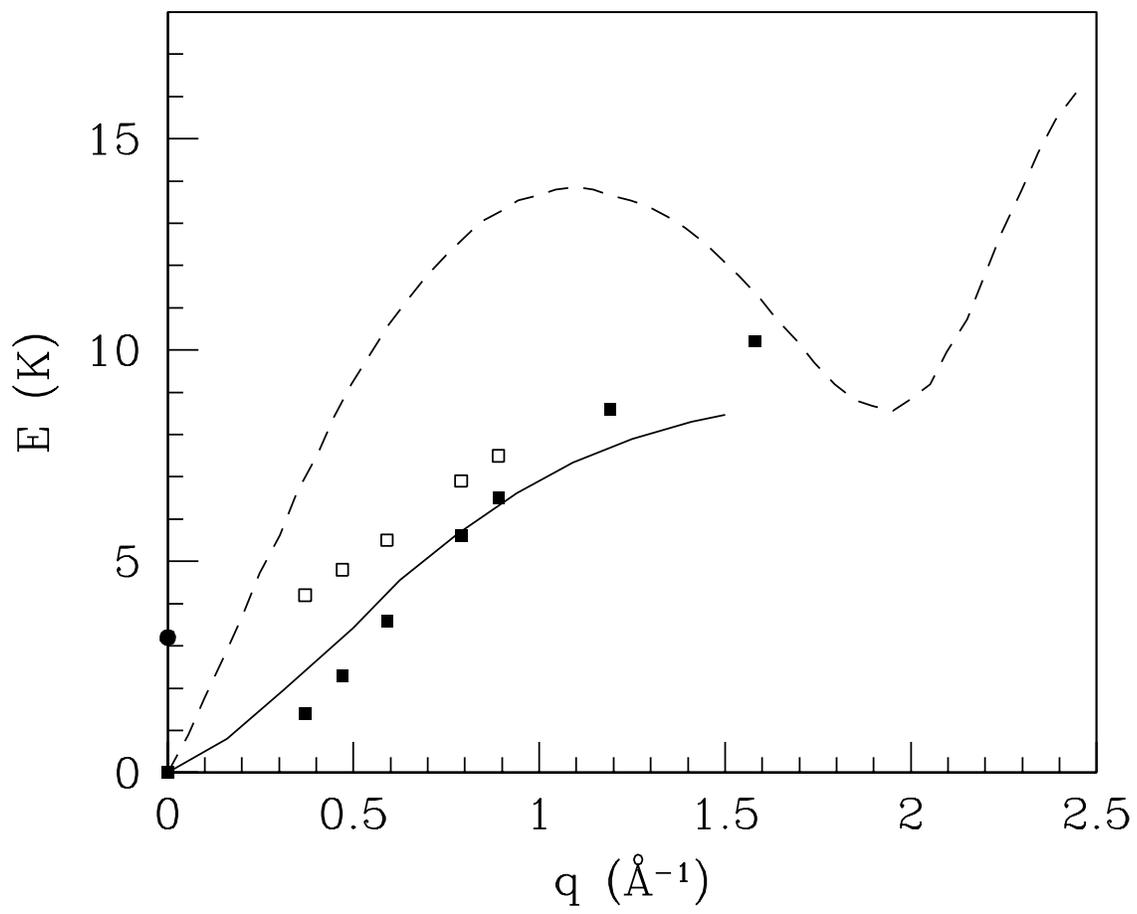
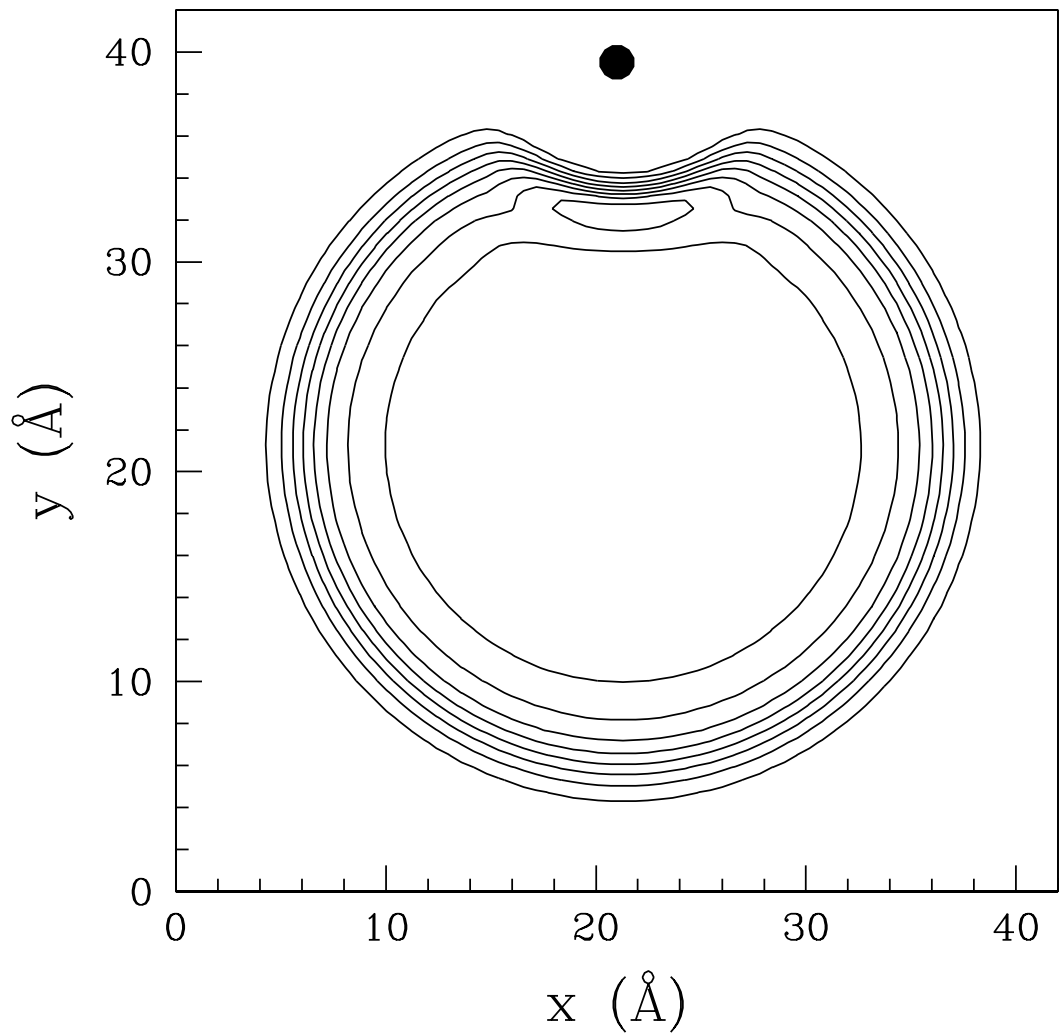
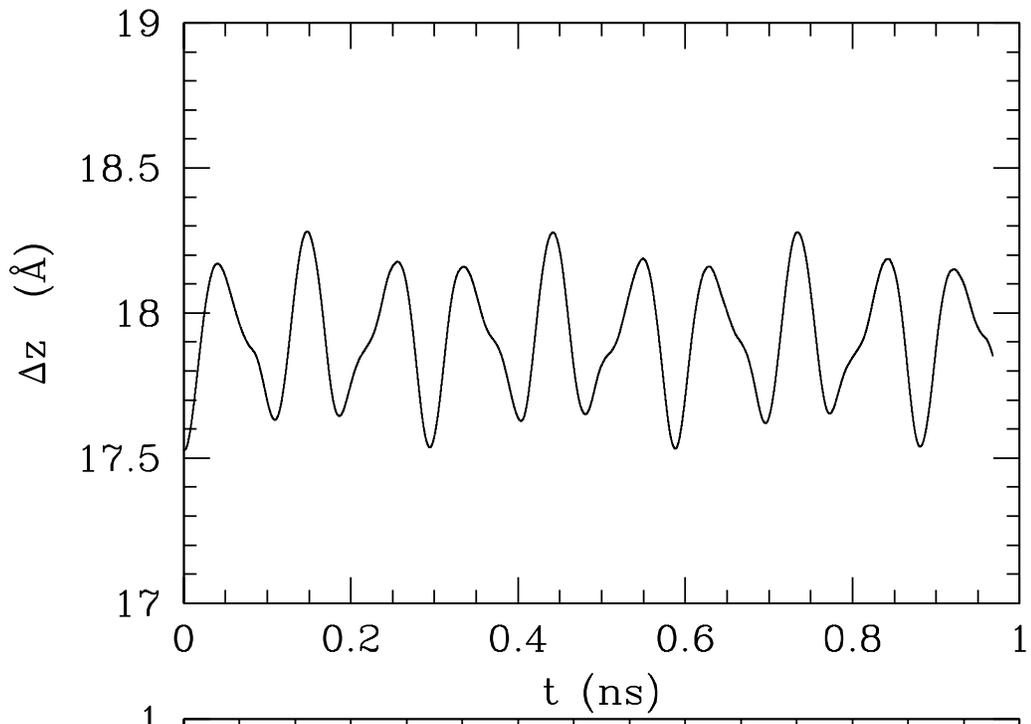
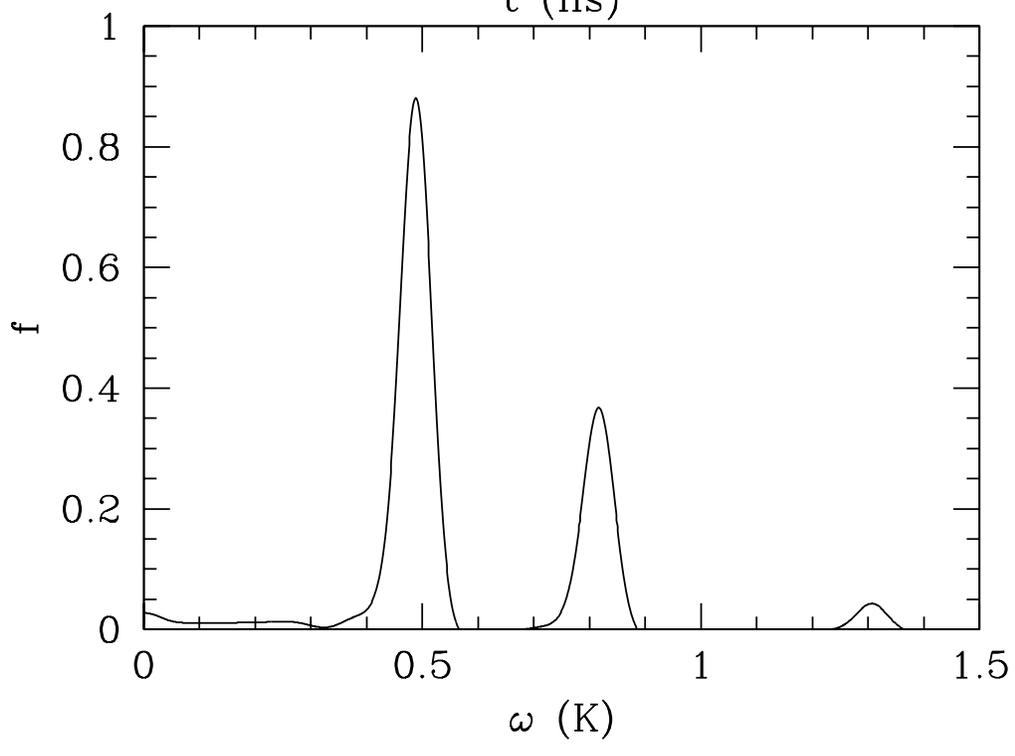
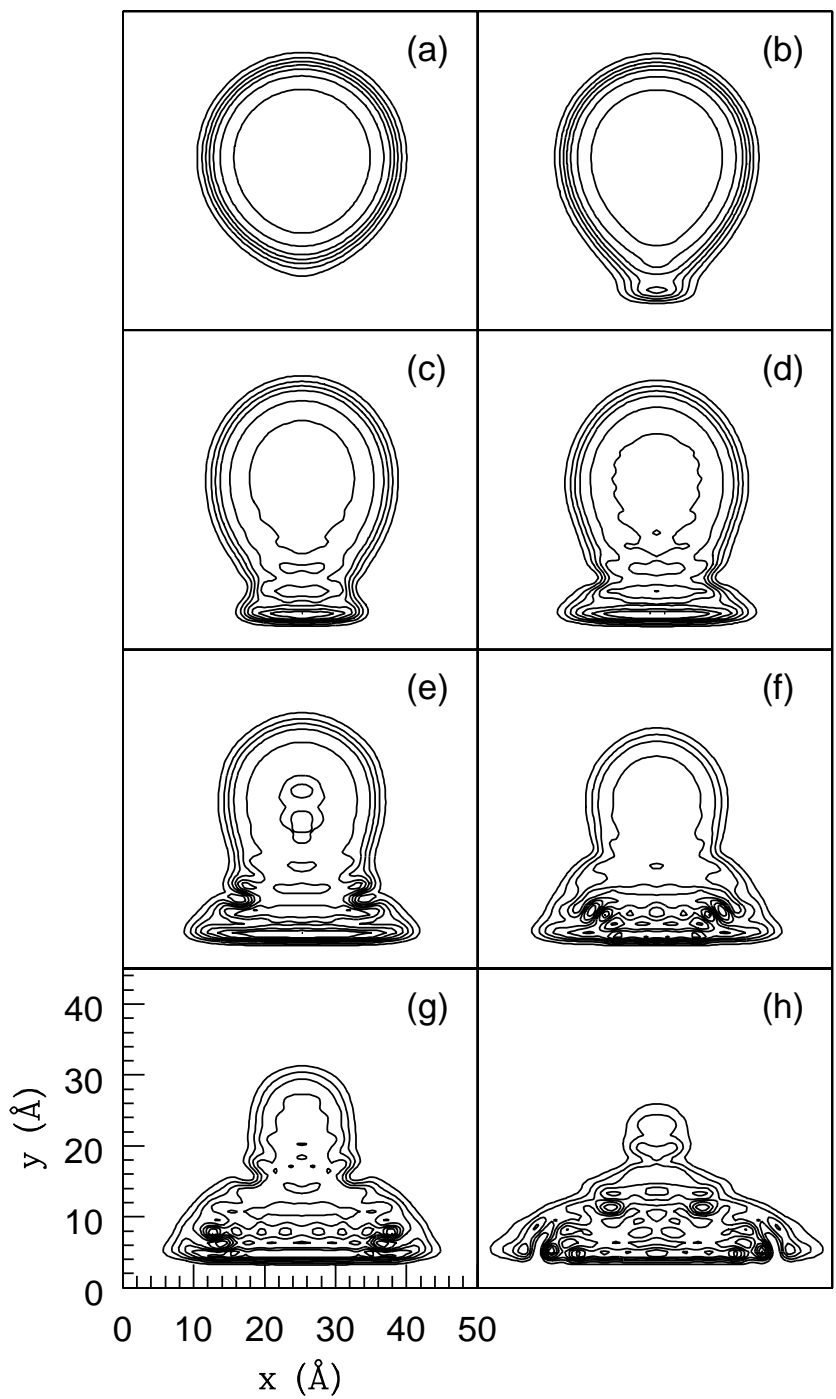
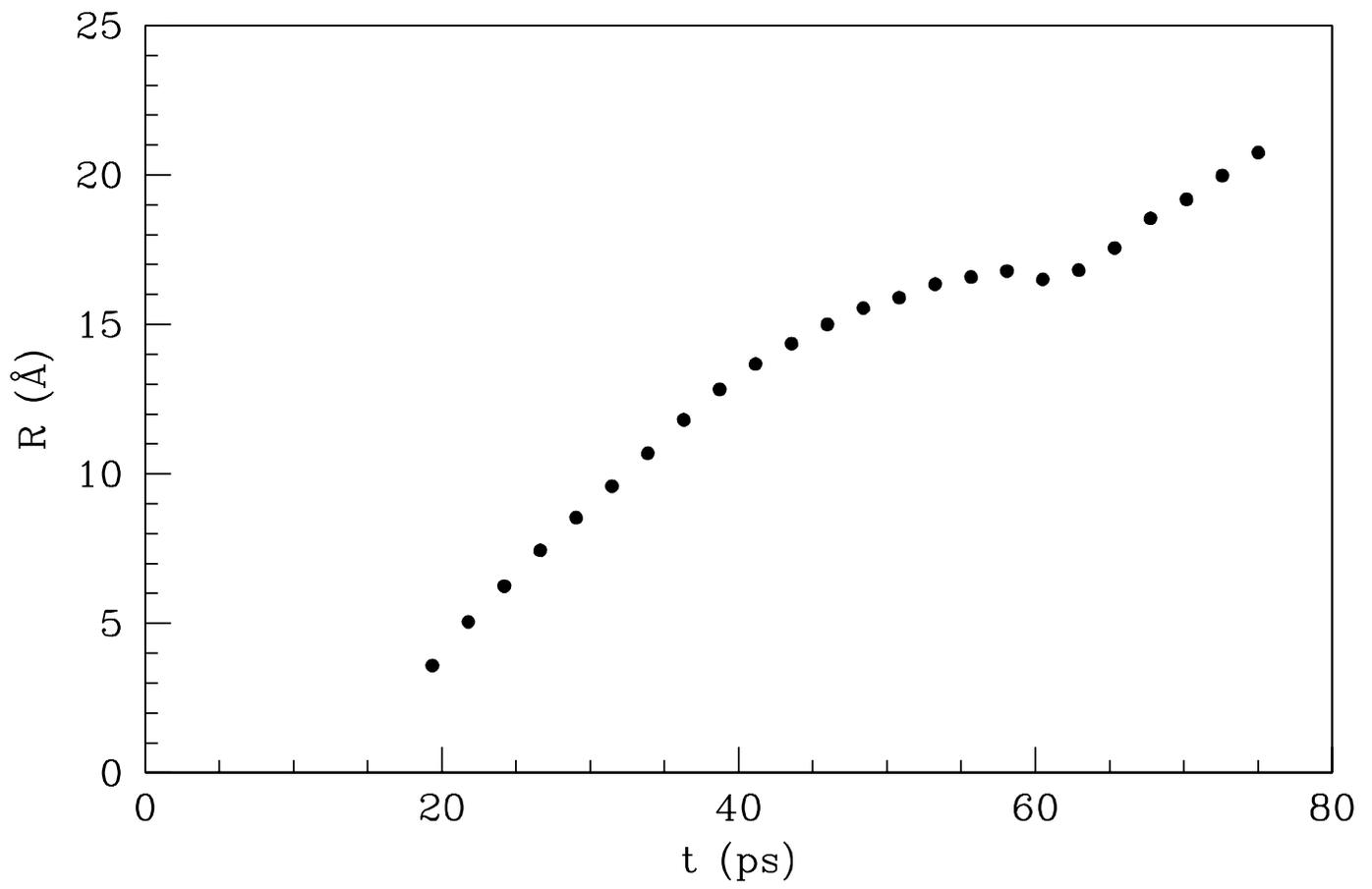
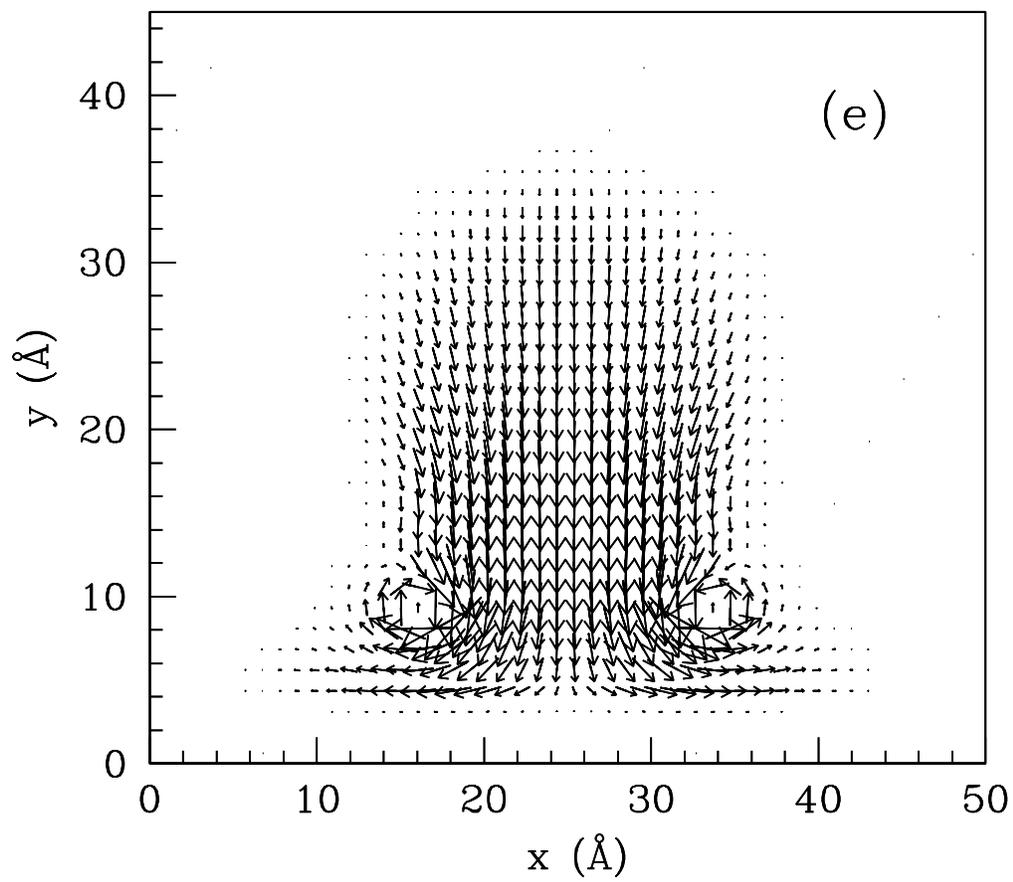
\end{document}